\documentclass[fleqn,onecolumn,10pt]{wlscirep}

\usepackage{graphicx}% Include figure files
\usepackage{dcolumn}% Align table columns on decimal point
\usepackage{bm}% bold math
\usepackage{multirow,caption}
\usepackage{amsmath}
\usepackage{hhline}
\usepackage{color}
\usepackage{xcolor}
\usepackage[normalem]{ulem}
\usepackage{array}
\usepackage{caption}
\usepackage{subcaption}
\usepackage[thinlines]{easytable}

\title{Structure and Evolution of Indian Physics Co-authorship Networks}

\author[1]{Chakresh Kumar Singh}
\author[1,*]{Shivakumar Jolad}
\affil[1]{Indian Institute of Technology Gandhinagar. Gandhinagar, 382355, INDIA }
\affil[*]{shiva.jolad@iitgn.ac.in}

\begin{abstract}
We trace the evolution of Indian physics community from 1919-2013 by analyzing the co-authorship network constructed from papers published by authors in India in American Physical Society  (APS) journals. We make inferences on India’s contribution to different branches of Physics and identify the most influential Indian physicists at different time periods. The relative contribution of India to global physics publication(research) and its variation across subfields of Physics is assessed. We extract the changing collaboration pattern of authors between Indian physicists through various network measures.  We study the evolution of Indian physics communities and trace the mean life and stationarity of communities by size in different APS journals. We map the transition of authors between communities of different sizes from 1970-2013, capturing their birth, growth, merger and collapse. We find that Indian-Foreign collaborations are increasing at a faster pace compared to the Indian-Indian. We observe that the degree distribution of Indian collaboration networks follows the power law, with distinct patterns between Physical Review A, B and E, and high energy physics journals Physical Review C and D, and  Physical Review Letters. In almost every measure, we observe strong structural differences between low-energy and high-energy physics journals.
\end{abstract}

\begin{document}

\flushbottom
\maketitle
\thispagestyle{empty}

%\noindent Please note: Abbreviations should be introduced at the first mention in the main text – no abbreviations lists. Suggested structure of main text (not enforced) is provided below.

\section*{Introduction}

Study of complex networks has revolutionized our understanding of several complex  phenomena observed in the fields such as biology, epidemiology, ecology,  neuroscience, sociology and economics \cite{albert2002statistical,newman2010networks,newman2003structure,dorogovtsev2002evolution}. Most real world networks are dynamic, as both the representative nodes and the relationship between the nodes keeps changing with time.  Analyzing  the evolution of such networks helps us to understand the structural and functional relationship between the variables of interest, and their dynamics  \cite{barrat2012dynamical, gross2008adaptive,skyrms2009dynamic}. In the last decade, network science community has studied various emergent phenomena in real world networks like  community structure \cite{newman2006modularity, fortunato2010community}, percolation transitions and cascading failures   \cite{motter2002cascade,crucitti2004model,li2012cascading}. 

Sociologists and economists have long realized the importance of social networks in the formation of social groups, political parties, professional bodies, trading networks, movie actors, and scientific communities.  Tracing the evolution of such networks help us to understand the dynamics of society, politics, trade, cinema and science. Scientific body of knowledge is formed by communities of scientists working together on problems of common interest.  The availability of metadata of papers,  authors,  references and citations, in scientific journals and archives drove a large body of work on understanding of scientific collaborations and their impact on science itself  \cite{newman2001structure,newman2001scientific,newman2001scientificII,hou2007structure}.

Study of scientific collaborations which reflect on the nature and growth of scientific disciplines dates back to the pre-digital era \cite{beaver1978studies,de1979studies,beaver1979studies,beaver2001reflections}. In recent decades,  digitization and documentation of scientific articles has enabled scientific community to build network of scientific collaborations and citations and trace their growth (see for eg: Dong \textit{et.al.} ~\cite{dong2017century}).   In the last two decades many studies have tried address questions on evolution of scientific disciplines, diffusion of knowledge across different fields, and emergent patterns reflected in communities. Recent studies have explored both the static \cite{newman2004coauthorship, newman2001structure,newman2001scientific,newman2001scientificII} and dynamic \cite{mali2012dynamic, gasko2016new,yan2009applying,servia2015evolution,newman2004best,lee2008evolution,lee2010complete,mcdowell2011evolution, pan2012evolution, enduri2015does},  co-authorship networks.

However, country-specific studies of co-authorship networks have been relatively rare, especially for developing countries like India.  Country-specific studies can give us insights into dynamics of scientific disciplines, science policies pursued and relative impact the country has made to specific disciplines. Previous studies on the Indian physics coauthorship studies have focused on the descriptive statistics of collaboration \cite{raina1995collaboration} and not on the network structure. In this work, we study the evolution of physics and its subfields  in India using co-authorship networks of articles published in APS journals having at least one author with  affiliation to an Indian University or an Indian academic or research Institution. We construct the co-authorship network of Indian physicists by using meta data of papers, authors and their affiliations from the APS(American Physical Society) Physical Review series of journals. We analyze the network structure and evolution in intervals of five years using the network centrality measures, community detection and evolution.  We tracked the communities including life cycle,  the affiliation of authors, and extracted influential authors in each journal at different time periods.

We address questions such as the contribution of Indian physicists to different branches of Physics, the collaboration of Indian physicists among themselves and with authors from outside India, identifying the most influential Indian physicists at different time periods.  We identify the network structure and trace the community dynamics in these journals from 1970-2013.  Our work sheds light into the structure of collaborations  Indian physics community, the relative contributions to different branches of Physics, community structure across different sub-fields,  collaboration with authors from other countries,  and helps us assess whether the characteristics are different from the global network.

\section*{Results}

\subsection*{Data }

%{\bf Developments from 1919-1969}: 

The Physical Review journals trace back to the late 1800s, with the first journal Physical Review (PR) started in 1893. The American Physical Society (APS) took over publishing in 1913 and continued publishing articles in PR till 1970. The Review of Modern Physics (RMP) started in 1929 (to date ) by APS, publishes reviews of current trends in research in fundamental physics and applications. Currently APS has divided its publications into broadly into six main journals Physical Review A to Physical Review E, (PRA, PRB, PRC, PRD, PRE) spanning different branches of Physics (PRA: Atomic and Molecular Physics; PRB: Condensed Matter and Material Physics; PRC: Nuclear Physics; PRD: Particle, Fields, Gravitation and Cosmology; PRE: Statistical, Non-linear, Biological and Soft Matter Physics),  and Physical Review Letters (PRL)  covering significant leads in all branches of physics. The PRL started in 1958, PRA to PRD were published from 1970, and in 1993 the PRE was added to publish research in statistical physics and allied fields. We have excluded APS publications in new journals which were started 2005 onwards.  In this study, we consider articles in APS publications from 1919-2013 with at least one author with an Indian affiliation. These articles may have authors entirely Indian or Indian and Foregn collaborations.

\subsection*{Descriptive Analysis }

\begin{figure*}[ht]
	\centering
	\includegraphics[scale=0.20]{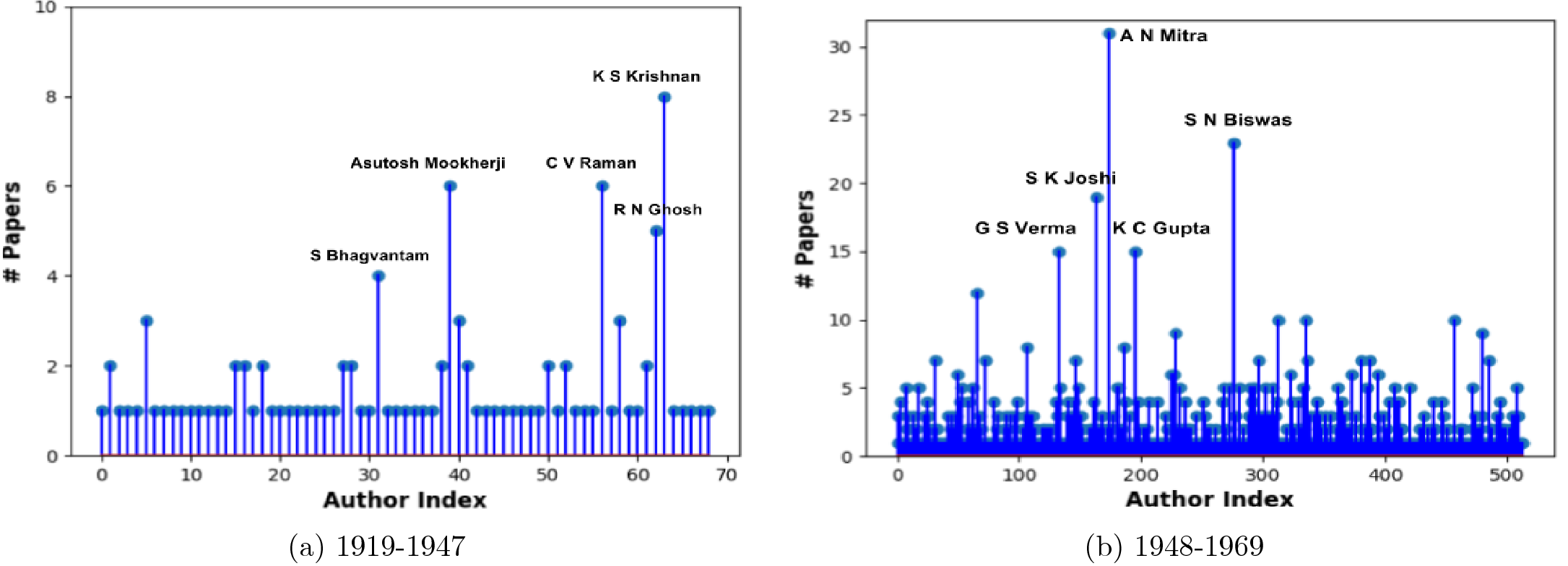}
	\caption{Indian authored papers in APS journals in Physical Review  and Review of Modern Physics  from 1919-1969, and Physical Review Letters from 1956-69. Panel represents (a) up to Indian independence 1947 and panel (b) post Independence period from 1948-1970. }
	\label{fig:bfr69}
\end{figure*}

Indian physicists have been publishing in APS journals since 1919, but their numbers were not significant till the 70's.  We give a brief summary of Indian authors papers in APS journal papers from 1919 to 1969 and focus more on the publications from 1970-2013. Indians contributed  a total of 695 papers between 1919-1969 in APS journals (Physical  Review and PRL). In the pre-independence (before August 1947) period many Indians were publishing papers in European journals which dominated scientific literature prior to world war II. Due to inaccessibility of data, we are excluding these journals from our analysis.   In Fig.$\ref{fig:bfr69}$ (a) and $\ref{fig:bfr69}$ (b), we show the number of papers published by Indian Authors pre-independence (1919-1947)  and post-independence (1948-1969) time periods up to 1969.  From 1919-47, a total of 85 papers were published by Indian authors.  Most of these papers had either one or two authors.   Many eminent Indian scientists like Sri C. V. Raman (Nobel Laureate, 1930), K S Krishnan , S Bhagavantam and Asutosh Mookherjee dominated the Indian Physics.  Post-independence, the number of publications by Indian increased substantially. In 21 years, Indians published 14,704 papers- many of them being two author papers.
%{\bf Papers from 1970-2013}: 

\begin{figure}[ht]  
	\centering
	\includegraphics[scale = 0.085]{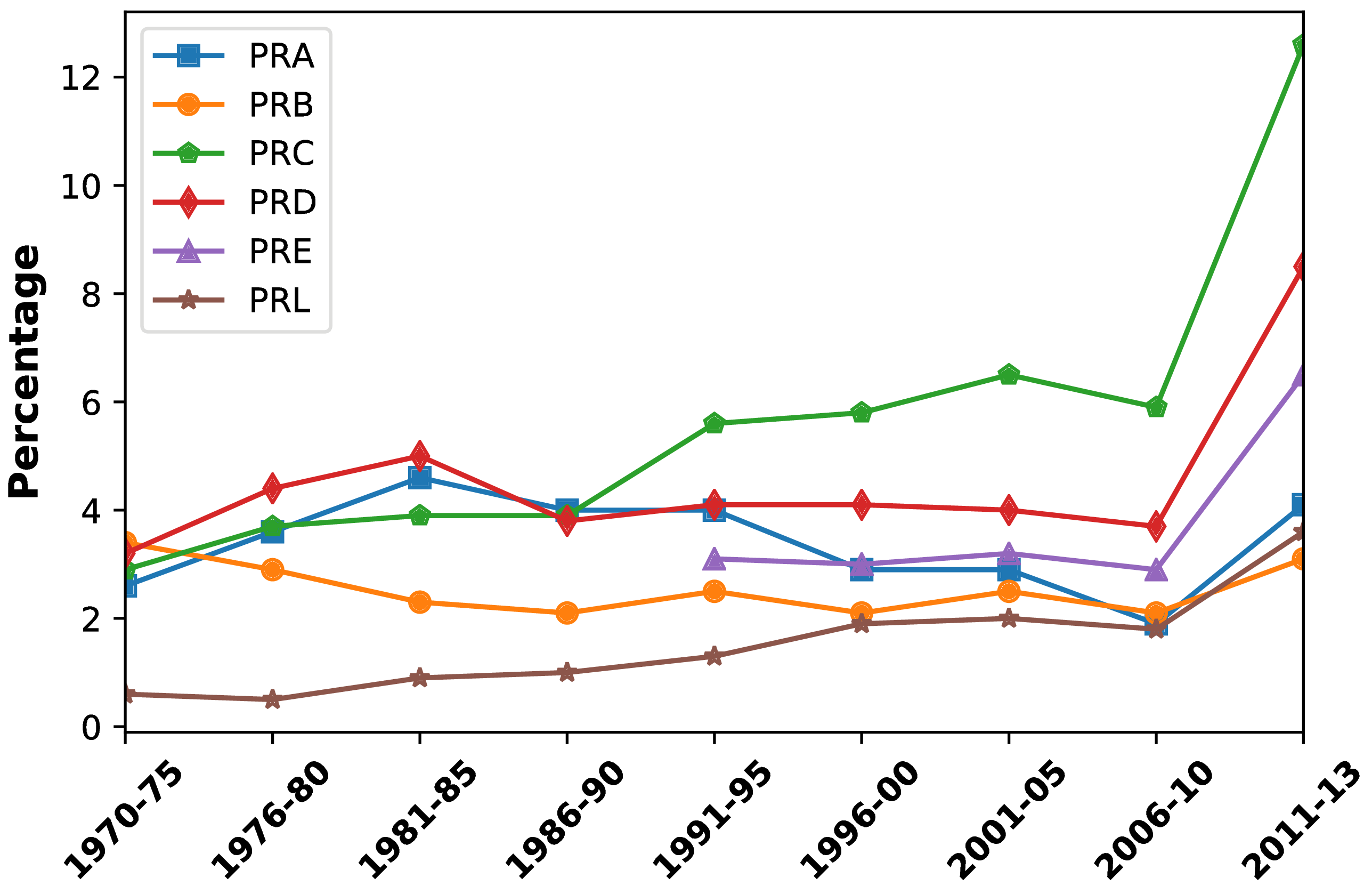}
	\caption{Percent of Indian papers in different APS journals from 1970-2013 in intervals of five years. }
	\label{fig:p_c}
\end{figure}

Between 1970-2013, a total of 486,440 papers were published in APS journals including PRA-E and PRL(see Table $\ref{indpapTot}$). Out of these 14,704 papers ($3.02\%$), had at least one author with Indian affiliation. In Fig. \ref{fig:p_c},  the percentage of papers with at least one Indian affiliation in different APS journals is shown. Indian contribution is highest in PRC (Nuclear and High energy Physics), and it rose from 2.9\% in 1970-75 to 12.6\% in 2011-13. In PRB, the contribution has been hovering around 2.5\%. In PRE, we see a marginal increase from 3.1\% in 1991-95 to 6.5\% in 2011-13. PRL (0.6\% to 3.6\%) and PRD (3.2\% to 8.5\%)  show concurrent rise in Indian contribution. We should be cautious while interpreting it as many high energy physics papers have large numbers of authors with multiple affiliations which raises the proportion of authors from a given country.

\begin{table}[ht]
	%\resizebox{0.35\textwidth}{!}{%
		\centering
		\begin{tabular}{lrrr}
			\hline
			&  \multicolumn{2}{c}{Papers}    &  \\ 
			Journals          \quad  & \quad Total\quad   & \quad Indian \quad  & \quad  Percent \\ \hline
			PRA      & 65,170        & 2,146          & 3.29\%  \\
			PRB      & 161,257       & 3,941          & 2.44\%  \\
			PRC      & 34,443        & 1,967          & 5.71\%  \\
			PRD      & 69,481        & 3,197          & 4.60\%  \\
			PRE      & 46,009        & 1,658          & 3.60\%  \\
			PRL      & 110,080       & 1,794          & 1.63\%  \\ \hline
			TOTAL    & 486,440       & 14,703         & 3.02\%  \\ \hline
		\end{tabular}%
	
	%}
	\caption{Contribution of Indian authored papers in different APS journals from 1970-2013. }
	\label{indpapTot}
\end{table}

\begin{figure*}[ht]
	\centering
	\includegraphics[scale=0.55]{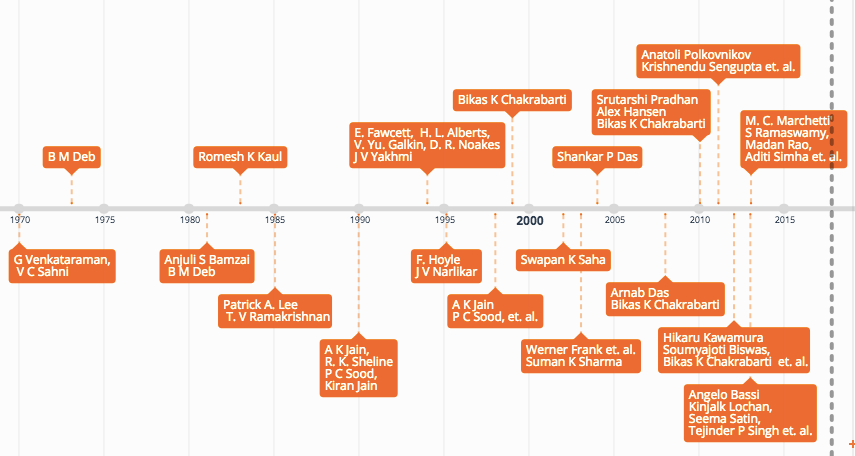}
	\caption{Publication of Indians in Review of Modern Physics from 1970-2013.}
	\label{fig:rmptime}
\end{figure*}

 We treat the Review of Modern Physics publications separately as only 33 publications by Indian authors can be found between 1919-2013. We highlight the Indian authored RMP papers from 1970-2013 as a timeline in Fig. $\ref{fig:rmptime}$. Topics covered in these review papers include almost all  branches of physics like condensed matter physics, cosmology, nuclear physics,  particle phenomenology, atomic physics, soft-matter and non-equilibrium statistical physics. We observe that RMP publications are becoming more frequent, and the number of multi-authored (three or more) papers are increasing. 

In Table \ref{stats}, we briefly describe the contributions by Indian authors in APS journals. PRB has the maximum number of authors ($N_{auth} \approx 3078$)  followed by PRD, PRL PRA and PRC respectively. $\langle P \rangle $ gives the mean number of papers published by an Indian author in the journal. In general, there are more authors per paper in PRL and PRC and PRD compared to other journals, which may be due to membership in large collaboration groups (for ex. D0, WA98 etc.) in high energy Physics. The mean active period (years) $\langle T \rangle $ column shows that  Indians on an average actively publish in APS journals for 4.4 years and at the rate of 3.88 papers in that period. The maximum mean average life span is low at 6.66 years. 

\begin{table}[ht]
	\centering
	%	\resizebox{0.30\textwidth}{!}{%
	\begin{tabular}{lrrrr}
		\hline
		&  \textbf{Authors}   & \textbf{Papers } & \textbf{Mean time} &  \textbf{Max papers} \\ 
		\textbf{Journals} & $N$ & $\langle P \rangle$  & $\langle T \rangle$  & $ M_p $ \\ \hline
		\textbf{PRA}      & 1608                           & 3.06            & 4.65                   & 211            \\    
		\textbf{PRB}      & 3078                           & 3.24            & 5.11                   & 94               \\    
		\textbf{PRC}      & 1562                           & 6.5             & 6.66                   & 102              \\  
		\textbf{PRD}      & 1933                           & 5.5             & 5.5                    & 170                \\ 
		\textbf{PRE}      & 1459                           & 2.4             & 3.38                   & 44                  \\
		\textbf{PRL}      & 1903                           & 6.9             & 4.2                    & 277                 \\
		\textbf{PR}      & 520                            & 2.1             & 3.8                    & 26                  \\
		\textbf{RMP}      & 42                             & 1.4             & 1.9                    & 5     \\     \hline         
	\end{tabular}%
	%	}
	\caption{Indian authors publications in different APS journals. $N$ is the number of distinct Indian authors, $\langle P \rangle$ is the average number of papers published,  $\langle T \rangle$ represent  the mean time in years for which an author is considered active and $ M_p $ is the maximum number of papers published by a author within the span of the dataset (refer to Appendix figures for more details).}
	\label{stats}
\end{table}

\subsection*{ Collaboration Coefficient }

Number of multi-authored paper by Indian physicists has steadily increased 1970's.   Collaboration Coefficient (CC) is a  simple statistical measure (non-network based) to assess the extent of multi-author papers and track its evolution (see Methods section for details). 

\begin{figure*}[ht]
	\centering
	\includegraphics[width=1.0\textwidth]{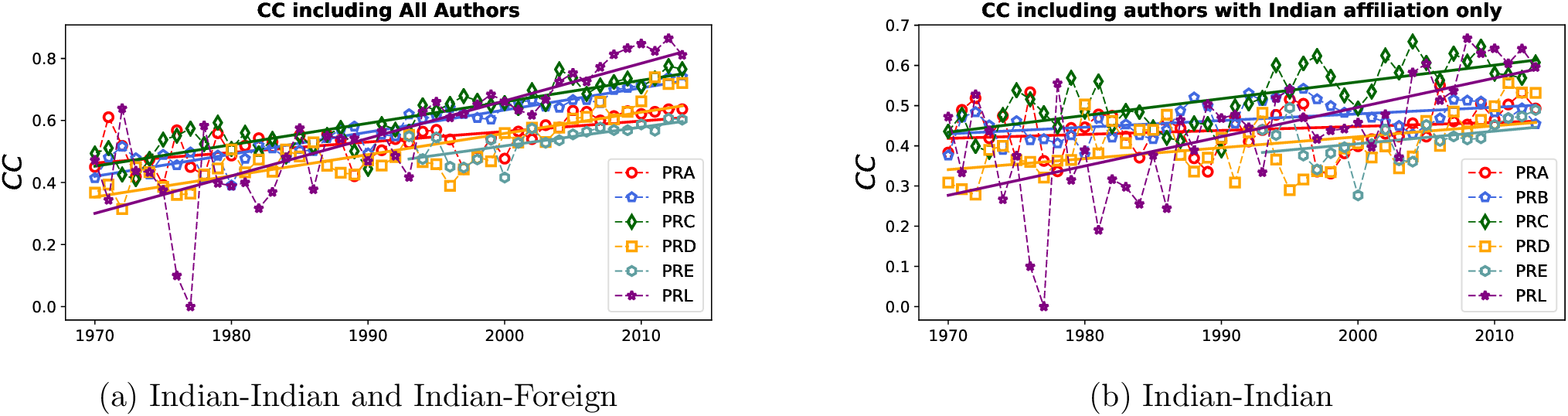}
	\caption{ Collaboration coefficient of papers (a)  having at least one author with Indian affiliation (I-IF)  (b)  only Indian authors (I-I) for Journals PRA-PRD, and PRL from 1970-2013. }
	\label{fig:cc}
\end{figure*}

 In Fig. $\ref{fig:cc}$ (a), CC has been estimated for all papers in Phys. Rev journals with at least one author from India. The CC for lies between 0.3 to 0.9.   A linear fit to the data shows that the slope varies considerably across the Journals, with  PRL showing the largest slope (0.012/year). Condensed Matter and Material Physics shows the largest increase in foreign collaborations.   In all Journals there is an increasing collaboration/multi-author papers over time.  In table \ref{cc_mt}, we show the slope of CC between Indian-Indians (I-I), Indian-Foreign (I-F), Indian with all (I-IF), for different journals. The collaboration of Indians with foreigners has been rising at a much faster rate than Indians with Indians. The PRL shows the highest increase collaborations in the I-F category (15.69). In Fig. $\ref{fig:cc}$ (b), we consider authors with Indian affiliations among all authors ignoring Foreign affiliations for calculating CC. We see that $CC$ ranged between 0.32-0.55 with a dip in PRL for the year 1976 indicating the presence of single authored papers (6 in number).   PRC and PRL having higher increase in rate of collaboration than PRA and PRB. PRA has papers with small groups, shows minimal gradient in CC.

%[XXX: 0.003/yr means 0.09 per decade or a rise of 0.27 in three decades. Are there significant difference in slope between PRA, PRB - versus PRC, PRD and PRL?. Why PRE has not been included? It started in 1993, lot of Indian authors are on it. Doesn't CC refer to papers ? What do you mean Including all authors and Only Indian authors. ]

\begin{table}[ht]
	\centering
%	\resizebox{0.48\textwidth}{!}{%
		\begin{tabular}{m{1.7cm} p{1.7cm} p{1.7cm} p{1.7cm}}
			\hline
			\multicolumn{4}{m{7.3cm}}{\textbf{Collaboration coefficient gradient ($10^{-3}/yr$)}} \\ 
			\textbf{Journal}           & \textbf{I-I } & \textbf{I-F} & \textbf{I-IF} \\ \hline
			\textbf{PRA} & 0.99                   & 2.91                   & 3.39                    \\ 
			\textbf{PRB} & 1.65                   & 8.16                   & 7.13                    \\ 
			\textbf{PRC} & 4.17                   & 7.42                   & 6.92                    \\ 
			\textbf{PRD} & 2.73                   & 8.4                    & 6.83                    \\ 
			\textbf{PRE} & 3                      & 4                      & 6                       \\ 
			\textbf{PRL} & 7.29                   & 15.69                  & 12.09                   \\ \hline    
		\end{tabular}%
%	}
	\caption{The slope of collaboration coefficient measure. I-I(Indian with Indian) considers only Indian authors in a paper, I-F(Indians with Foreign) only foreign authors and I-IF(Indian with Indian and Foreigners). }
	\label{cc_mt}
\end{table}

\begin{figure*}[ht]
	\centering
	\includegraphics[width=0.8\textwidth]{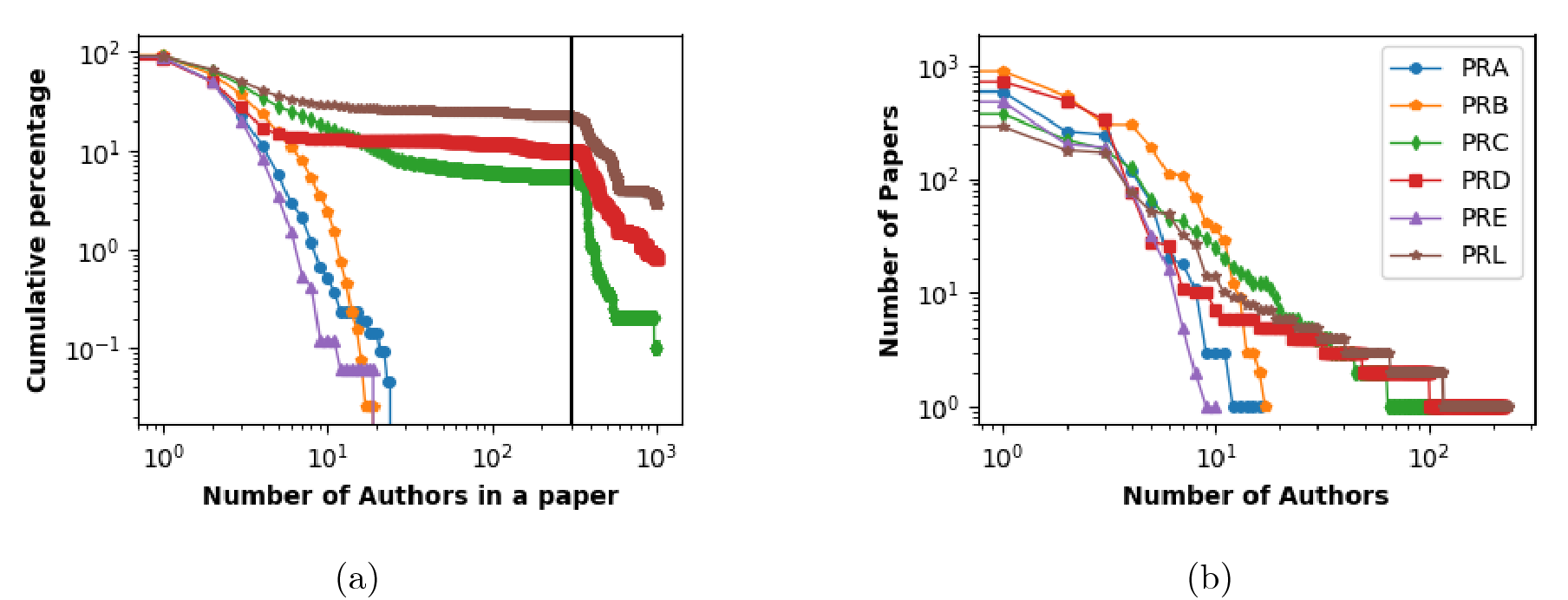}
	\caption{(a) The cumulative distribution of number of authors in a paper. Black line (at 300) is drawn  where there is a sharp  dip in the fraction of papers for PRC, D and L  (b) the distribution of the number of papers vs. number of authors in a paper. }
	\label{fig:c_auth}
\end{figure*}  

\subsection*{Co-authorship network}
The collaboration coefficient measure is a descriptive measure and does not capture the complex network structure. In this section, we examine the distribution of authors and paper in each of the Phys Rev journals and describe the methods used for constructing co-authorship networks. In Fig. \ref{fig:c_auth} (a), the cumulative percentage of the number of authors in a paper, and in Fig. \ref{fig:c_auth}, the number of papers v/s number of authors in a paper for journals PRA-PRE, and PRL are shown.  We see district difference between Phys Rev A, B and E, and high energy physics journal Phys Rev C and D and PRL. In PRA, PRB and PRE, the distribution authors rapidly fall off after reaching order 10. They represent research done in small groups (theory, computational and low energy experimental physics). Whereas PRC and PRD cover high energy physics papers with a large number of authors of the order  $100$ to $1000$'s. In these areas, it is fairly common to have large collaborative research groups across continents such as LHC and LIGO.   Even though such papers form a small percent of the total (see Fig. \ref{fig:c_auth} (a)),  they have large impact on the network as they form dense cliques  (of size $\frac{m(m-1)}{2}$ edges, with m is the number of authors in the paper) in the network. Such large cliques do not indicate collaboration between every pair of scientists in the paper based on mutual interaction. We set a natural breaking point  (for PRC, PRD, and PRL) after 300 as a line in Fig. \ref{fig:c_auth} a, to eliminate such collaborations which are unrealistic and skew the network measures. Choosing this number as a cut-off is an optimal trade-off between collaborators with interpersonal and professional relationships vs association through membership in large collaboration groups.

\begin{table*}[ht]
	\centering
	%\resizebox{0.7\textwidth}{!}{%
		\begin{tabular}{|c|rr|rr|rr|rr|rr|rr|}
			% 	\begin{tabular}{c|cccccccccccc}
			\hline
			% 		\multicolumn{13}{c}{Graph size of APS Indian Physics networks with Indian and Foreign authors} \\ \hline
			& \multicolumn{2}{c|}{~~~\textbf{PRA}~~~}      & \multicolumn{2}{c|}{~~~\textbf{PRB}~~~}        & \multicolumn{2}{c|}{~~~\textbf{PRC}~~~}       & \multicolumn{2}{c|}{~~~\textbf{PRD}~~~}       & \multicolumn{2}{c|}{~~~\textbf{PRE}~~~}       & \multicolumn{2}{c|}{~~~\textbf{PRL}~~~}         \\ 
			& I-IF & I-I   & I-IF & I-I     & I-IF & I-I  & I-IF & I-I & I-IF & I-I & I-IF & I-I \\ \hline
			\textbf{1970}          & 17 & 15                 & 50 & 45                 & 28 & 25                 & 54 & 49                 & NA & NA                     & 13 & 13                 \\ 
			\textbf{1980}          & 59 & 54                 & 64 & 63                 & 61 & 59                 & 63 & 62                 & NA & NA                     & 18 & 18                 \\ 
			\textbf{1990}          & 118 & 104               & 149 & 128               & 72 & 65                 & 147 & 140               & 92 & 76           & 139 & 139               \\ 
			\textbf{2000}          & 87 & 72                 & 329 & 205               & 161 & 109               & 656 & 149               & 134 & 98                & 795 & 154               \\  	
			\textbf{2010}          & 206 & 143               & 585 & 304               & 1453 & 304              & 4670 & 365              & 309 & 228               & 5053 & 271              \\ \hline
		\end{tabular}
	%}
	\caption{Graph size of APS Indian Physics networks with Indian and Foreign authors. I-IF is the network with Indian and Foreign authors, I-I is the network with only Indian authors. }
	\label{graph_evol}
\end{table*}

\subsection*{Network Characteristics and Growth}

Coauthorship networks provide an opportunity to study the emergence of communities and their evolution.Co-authorship networks at a global level been studied in multiple fields such as Computer Science, Biological and Physical sciences \cite{newman2001structure,newman2001scientific,newman2001scientificII, newman2004coauthorship}. These studies reveal that co-authorship these networks exhibit small-world characteristics with high clustering and follows a power-law distribution with exponential cutoff \cite{mali2012dynamic}. Coauthorship networks are dominated by preferential attachment and formation of triads, thus increasing clustering, as the network grows with time giving rise to the power-law degree distributions followed by these networks \cite{newman2001clustering, lee2008evolution, lee2010complete, mcdowell2011evolution}. Some attempts at modeling such networks based on  observations of formation of triads and preferential attachment have been made \cite{lee2008evolution, lee2010complete}.  In this section, we provide descriptive statistics of Indian coauthorship network,  characterize the growth of network from 1970-2013 and extract the influential authors in this time period. 

The co-authorship network of Indian physicists is weighted, undirected and dynamic. In Table \ref{graph_evol},we show a sample of the network growth (graph size from 1970 to 2010)  with time for all the journals, counting only Indians and Indian-Foreign authors together.  The graph size with foreign authors becomes large in high energy physics journal and PRL.

\begin{figure*}[ht]
	\centering
	\includegraphics[width=0.9\textwidth]{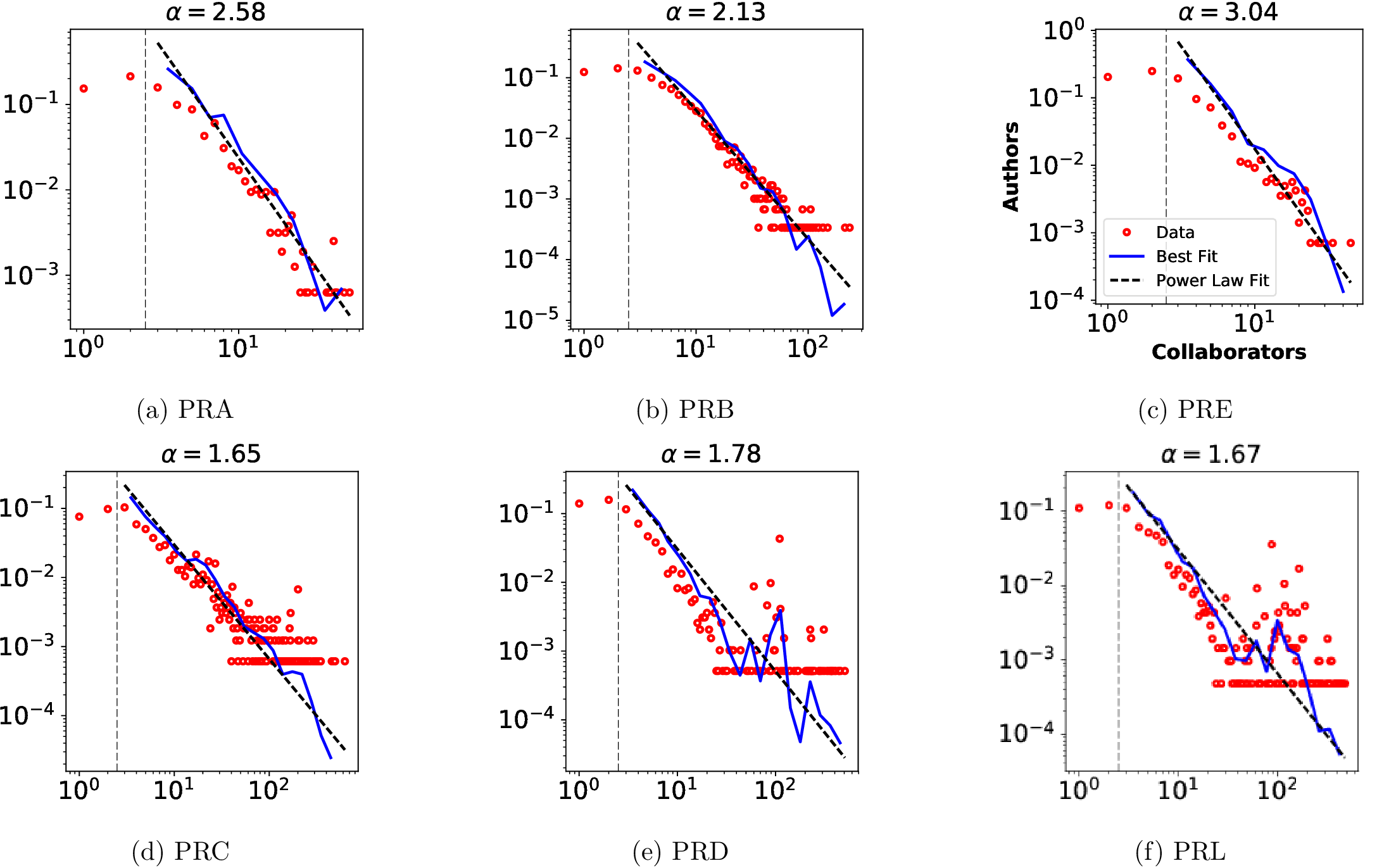}
	\caption{Distribution of collaborators of authors for APS journals Phys Rev A-E and Letters. Red dots represent the data,  blue curve is the binned data using power law package \cite{alstott2014powerlaw}, and black dotted line shows the power law fit to binned with  $x_{min}=3$ (at the dotted vertical line), and blue curve represent the best data . }
	\label{fig:auth_colab_dist}
	
\end{figure*}

Distribution of total collaborations of authors for major APS journals is shown in Fig. $\ref{fig:auth_colab_dist}$ panels (a-f). The upper bound for the number collaborators is set to 300 as described in the earlier section,  keeping in view of the large papers in Phys Rev C, D and Letters.  The binned data (blue) is fitted to a power law distribution $P(k) \propto k^{-\alpha}$ (using python package powerlaw ~\cite{alstott2012powerlaw} ) as shown in (black dotted line) by setting $x_{min}=3$, and the exponent $\alpha$ is shown in each panel. 

The exponent for Phys Rev A, B and E lies between 2 to 3.04, implying that they have a well defined mean degree. Whereas for PRC, PRD and PRL journals, the exponents lie between 1.5 to 2.0 indicating a long tail and a divergent mean (in the infinite limit). This is mainly due to articles by large collaborative experimental groups across the world which typically have authors ranging between hundreds to thousands.

\begin{figure*}[ht]
	\centering
	\includegraphics[width=1.0\textwidth]{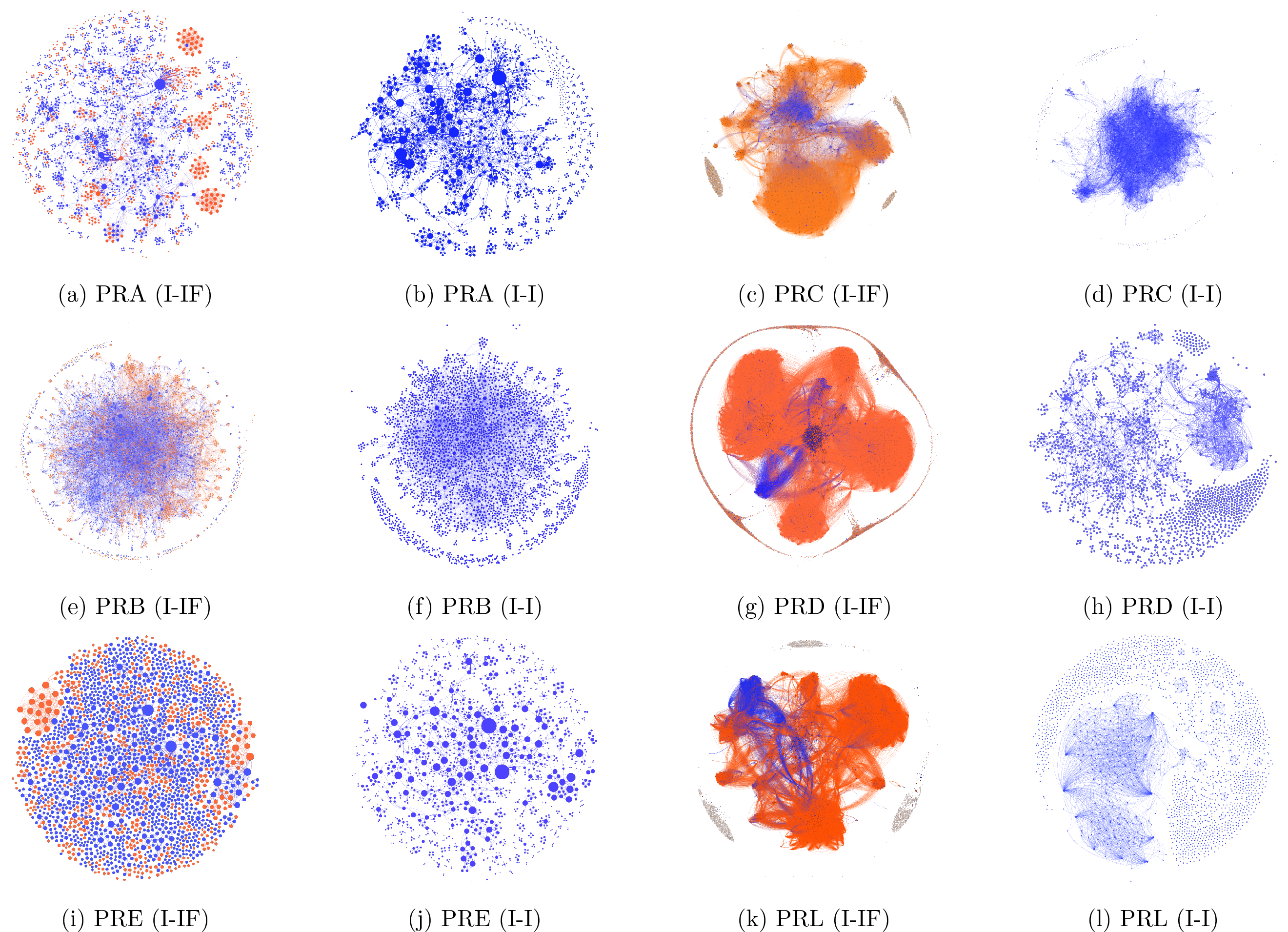}	
	\caption{Indian Physics co-authorship Network for journals PRA -L  from 1970-2013. I-IF represents Indians with Indian and foreign authors, I-I represents network of Indians with Indian authors only.  Indians are represented with Blue while foreign authors with orange color.  Network statistics are mentioned in Table. \ref{graph_stats}}
	\label{fig:complete_ntwrk_1}
\end{figure*}

\begin{table*}[ht]
	\centering
	%\resizebox{0.8\textwidth}{!}{
		\begin{tabular}{|c|cc|cc|cc|cc|cc|cc|cc|cc}
			\hline
			& \multicolumn{2}{c|}{PRA}      & \multicolumn{2}{c|}{PRB}        & \multicolumn{2}{c|}{PRE}       & \multicolumn{2}{c|}{PRD}       & \multicolumn{2}{c|}{PRC}       & \multicolumn{2}{c|}{PRL}         \\ 
			& ~~I-IF~~ & ~~I-I~~   & ~~I-IF~~ & ~~I-I~~   & ~~I-IF~~ & ~~I-I~~    & ~~I-IF~~ & ~~I-I~~ & ~~I-IF~~ & ~~I-I~~ & ~~I-IF~~ & ~~I-I~~ \\ \hline
			$\langle k \rangle$ & 5.3 & 4.1  & 8.3 & 6.2   & 3.9 & 3.1    & 33.07 & 8.4 & 31 & 7.14   & 27.17 & 13.96 \\ 
			$ \langle C \rangle$ & 0.8 & 0.76 & 0.82 & 0.74 & 0.82 & 0.77 & 0.82 & 0.72 & 0.85 &0.76 & 0.9 & 0.84    \\ 
			$\langle l \rangle$ & 6.4 & 6.31 & 4.5 & 4.5   & 6.75 & 6.64 & 4.37 & 6.49 & 3.26 & 3.65 & 3.98 & 5.08 \\ \hline  
		\end{tabular}
	%}
	\caption{Network statistics- average degree $\langle k \rangle$, clustering coefficient $\langle C \rangle$ and mean path length  $\langle l \rangle$  of Indian-Indian and Indian-foreign coauthorship network for APS journals from 1970-2013.}
	\label{graph_stats}
\end{table*}

\subsection*{Network measures}
In Fig. $\ref{fig:complete_ntwrk_1}$, we show the complete networks for PRA-PRL journals from (1970-2013), with all authors (I-IF) and with only Indian authors (I-I).
We see clear structural difference between these two sets of networks (PRA, B, E; and PRC,D and L). PRA has small clusters both for I-IF and I-I networks. Removal of foreign does not significantly change the network topology. We do find some clusters with mostly foreign and a single Indian authors possibly indicating long distance collaboration.  The I-IF network of PRC is dominated by few large clusters with mostly foreign nodes. Removal of foreign nodes makes the network sparse, but still dominated one Giant cluster.  While in PRD, the I-IF network is dense with large cliques dominated by foreign nodes. Removal of the Foreign nodes in the I-F makes the network relative sparse, with several disconnected nodes. The PRB and PRE networks have similar structure as PRA, while PRL has greater resemblance with PRD. 

In Table $\ref{graph_stats}$, we notice these topological differences through network measures average degree $\langle k \rangle$, clustering coefficient $\langle C \rangle$ and mean path length $\langle l \rangle$. The average degree does not change much for PRA, B and E. Whereas for PRC (31 to 7.4) , D (33.04 to 8.4) and L (27.17 to 13.96) , there are large changes in average degree. Path length also remains roughly same PRE, B and E for both $I-IF$ and $I-I$. The path increases significantly when we remove foreign authors from PRC (3.26 to 3.65), D(4.37 to 6.49) and L (3.98 to 5.08), highlighting the contributions of foreign collaborations in these journals. 

\subsection*{Influential Authors}

\begin{figure}[ht]
	\centering
	\includegraphics[width=0.35\textwidth]{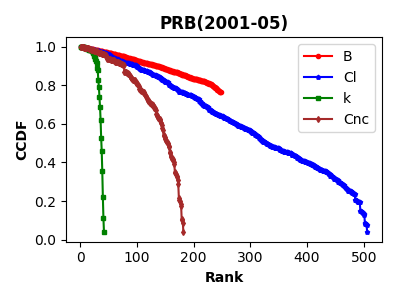}
	\caption{The Complementary Cumulative Rank  Distribution (CCRD) based on four centrality measures described i text for APS Indian co-authorship network of PRB between 2001-05. }
	\label{fig:cm_ranking}
\end{figure} 

Growth of scientific disciplines is largely driven by individuals who make significant contributions to the field by leading collaborative research groups. We have tried to identify Indian authors who have maximal capability to diffuse the knowledge across the research community by using network centrality measures. In complex networks, different centrality measures such as degree centrality, betweenness centrality, closeness centrality, and neighborhood coreness  centrality are used to extract the importance of nodes\cite{newman2004best, servia2015evolution, yan2009applying}. Relative value of these measures helps in ranking nodes based on their topological position within the network.   Each measure has distinct interpretation of influence (see methods section for details). Measures that award maximum rank dispersion for the nodes are usually preferred for ranking. As an illustration, in Fig. \ref{fig:cm_ranking}  we show the complementary cumulative rank distribution ($CCDF(r)=P(X\ge r)$,  fraction of nodes with rank $r$ or more)  based on betweenness $B$, closeness $Cl$, degree $k$ and neighborhood coreness $C_{nc}$ for PRB co-authorship network between 2001-05. 

 In Table.  \ref{influ_table}, we list the top  Indian authors who have been influential in four centrality measures referred above. We rank the authors based on above mentioned measures separately,. Intersection of sets constructed from the top $20\%$ authors from each ranking is chosen and Indian authors extracted from this set are considered the most influential in their respective time periods.  We get different set of influential authors for different time periods, barring few exceptions. This is correlated with the fact that mean active time period of authors in these journals is between 3-7 years as shown in Table \ref{indpapTot}. Very rarely we find authors being influential for long time period. Influential authors in high energy physics are mostly experimentalists working in large groups. The choice of threshold  influentials the size and characteristics of networks.

\begin{table*}[ht]
	\centering
	\resizebox{\textwidth}{!}{%
		\begin{tabular}{|l|l|l|l|l|l|l|l|l|}
			\hline
			\multicolumn{9}{|c|}{\Large Influential Indian authors by various centrality measures}                                                                                                                                                         \\[8pt]
			 \textbf{1970- 75}    & \textbf{1976-80}    & \textbf{1981-85}      & \textbf{1986-90}        & \textbf{1991-95}       & \textbf{1996-2000}    & \textbf{2001-05}         & \textbf{2006-09}  & \textbf{2010-13}      \\ [5pt] \hline
			\multicolumn{9}{|c|}{\textbf{PRA}}                                                                                                                                                                                  \\ \hline
			S. C. Mukherjee      & S. C. Mukherjee    & B. Sanjeevaiah       & V. Lakshminarayana     & M. M. Panja           & M. B. Kurup          & D. G. Kanhere             & H. Singhal         & H. Singhal             \\
			K. N. Pathak         & B. C. Saha         & A. N. Tripathi       & V. Gopalakrishna       & R. R. Puri            & K. G. Prasad         & Anil Kumar                & R. K. Chaudhuri    & J. A. Chakera          \\
			R. Prasad            & M. K. Srivastava   & S. C. Mukherjee      & R. Srivastava          & M. Azam               & M. Sarkar            & P. C. Deshmukh            & D. Mukherjee       & L. C. Tribedi          \\
			K. Roy               & S. K. Sur          & Shyamal Datta        & S. Bhuloka Reddy       & G. S. Agarwal         & K. Vijayalakshmi     &                           & R. A. Khan         & T. S. Mahesh           \\
			S. C. Deorani        & G. S. Agarwal      & S. K. Sur            & V. Radha Krishna Murty & C. L. Mehta           & H. C. Padhi          &                           & Rajat K. Chaudhuri & Ujjwal Sen             \\ \hline
			\multicolumn{9}{|c|}{\textbf{PRB}}                                                                                                                                                                                  \\ \hline
			Chanchal K. Majumdar & T. K. Saxena       & A. N. Basu           & S. K. Dhar             & S. K. Dhar            & U. V. Varadaraju     & S. K. Dhar                & B. L. Ahuja        & Chandrabhas Narayana   \\
			G. Rama Rao          & M. P. Verma        & Vijay A. Singh       & B. D. Paladin          & R. Vijayaraghavan     & B. K. Godwal         & A. K. Tyagi               & S. K. Dhar         & S. Singh               \\
			M. P. Verma          & K. N. Pathak       & E. V. Sampathkumaran & S. M. Kanetkar         & Z. Hossain            & R. Vijayaraghavan    & S. Rayaprol               & S. S. Banerjee     & S. K. Dhar             \\
			S. K. Joshi          & O. P. Sharma       & S. Prakash           & L. C. Gupta            & H. R. Krishnamurthy   & Z. Hossain           & Amish G. Joshi            & Sugata Ray         & Sugata Ray             \\
			C. N. R. Rao         & S. K. Joshi        & B. D. Padalia        & J. V. Yakhmi           & A. S. Tamhane         & H. R. Krishnamurthy  & A. K. Sood                & Ranjan Mittal      & A. K. Tyagi            \\ \hline
			\multicolumn{9}{|c|}{\textbf{PRC}}                                                                                                                                                                                  \\ \hline
			M. K. Mehta          & S. B. Manohar      & V. K. Mittal         & M. Saha                & A. Chatterjee         & A. Chatterjee        & D. S. Mukhopadhyay        & B. J. Roy          & B. J. Roy              \\
			S. L. Gupta          & Ashok Kumar        & P. Singh             & Y. P. Viyogi           & Y. P. Viyogi          & S. K. Gupta          & S. Chattopadhyay          & R. Palit           & R. Palit               \\
			S. Kailas            & S. Kailas          & Gulzar Singh         & A. Goswami             & C. Bhattacharya       & R. Dutt              & M. R. Dutta Majumdar      & V. Jha             & V. Jha                 \\
			S. C. Pancholi       & Satya Prakash      & Satya Prakash        &                        & P. Singh              & K. Kar               & Y. P. Viyogi              & S. Mukhopadhyay    & S. Santra              \\
			& Y. P. Viyogi       & M. V. Ramaniah       &                        & S. Kailas             & C. Das               & N. K. Rao                 & A. Saxena          & S. Mukhopadhyay        \\ \hline
			\multicolumn{9}{|c|}{\textbf{PRD}}                                                                                                                                                                                  \\ \hline
			N. Panchapakesan     & N. S. Arya         & G. Singh             & I. S. Mittra           & J. B. Singh           & D. P. Roy            & Probir Roy                & Probir Roy         & B. Bhuyan              \\
			G. Rajasekaran       & T. R. Govindarajan & N. Mukunda           & M. Kaur                & J. M. Kohli           & H. Mendez            & M. Sami                   & Nilmani Mathur     & Anirban Kundu          \\
			K. Datta             & J. N. Misra        & Ashok Goyal          & N. K. Rao              & S. K. Gupta           & V. Kapoor            & Ashok Goyal               & Ritesh K. Singh    & Monoranjan Guchait     \\
			A. N. Mitra          & A. Mozumder        & S. N. Biswas         & V. K. Gupta            & N. K. Rao             & B. Choudhary         & Soumitra SenGupta         & J. B. Singh        & R. Sinha               \\
			L. K. Pandit         & S. Roy             & V. K. Gupta          &                        & I. S. Mittra          &                      & Biswarup Mukhopadhyaya    & Manmohan Gupta     & Debajyoti Choudhury    \\ \hline
			\multicolumn{9}{|c|}{\textbf{PRE}}                                                                                                                                                                                  \\ \hline
			\multicolumn{4}{|c|}{}                      & S. R. Sharma          & Arnab Majumdar       & Moumita Das               & S. Kumar           & Subir K. Das           \\
		\multicolumn{4}{|c|}{}                         & M. Saha               & Debashish Chowdhury  & S. Mitra                  & Sudeshna Sinha     & Sanjay Puri            \\
		\multicolumn{4}{|c|}{}                         & Ramakrishna Ramaswamy & P. S. Goyal          & S. S. Manna               & M. Lakshmanan      & Dipankar Bandyopadhyay \\
			\multicolumn{4}{|c|}{}                          & A. Sen                & S. Lakshmibala       & C. Dasgupta               & A. K. Sood         & M. Lakshmanan          \\
			\multicolumn{4}{|c|}{}                          & P. Nandy              & Sriram Ramaswamy     & Ramakrishna Ramaswamy     & D. V. Senthilkumar & B. Kundu               \\ \hline
%			&                    &                      &                        &                       &                      &                           &                    &                        \\ \hline
			\multicolumn{9}{|c|}{\textbf{PRL}}                                                                                                                                                                                  \\ \hline
			Virendra Singh       & R. Ramachandran    & I. S. Mittra         & I. S. Mittra           & R. Raniwala           & D. S. Mukhopadhyay   & P. K. Kaw                 & Bijaya K. Sahoo    & S. P. Behera           \\
			K. Govindarajan      & A. N. Mitra        & R. Joseph            & S. K. Badyal           & S. Kachroo            & S. K. Badyal         & A. Kumar                  & S. P. Singh        & A. D. Ayangeakaa       \\
			P. Chandra Sekhar    &                    & S. K. Tuli           & M. M. Aggarwal         &                       & M. D. Trivedi        & S. S. Sambyal             & T. Senthil         & L. C. Tribedi          \\
			&                    & Y. Prakash           & N. K. Rao              &                       & S. S. Sambyal        & P. Raychaudhuri           & S. S. Sambyal      & G. Ravindra Kumar      \\
			&                    & S. Satti             & V. K. Gupta            &                       & M. R. Dutta Majumdar & Somendra M. Bhattacharjee & P. K. Kaw          & S. S. Ghugre     \\ \hline     
		\end{tabular}%
	}
	\caption{Top influential Indian authors in APS Indian coauthorship network in five year intervals from 1970-2013. These  authors have been chosen based on their consistently top ranking in four centrality measures: Betweenness centrality, Closeness centrality, degree centrality and neighbourhood core centrality. }
	\label{influ_table}
\end{table*}

\subsection*{Communities in Indian Physics Co-authorship Network}

\begin{figure*}[ht]
	\centering
	\includegraphics[width=0.90\textwidth]{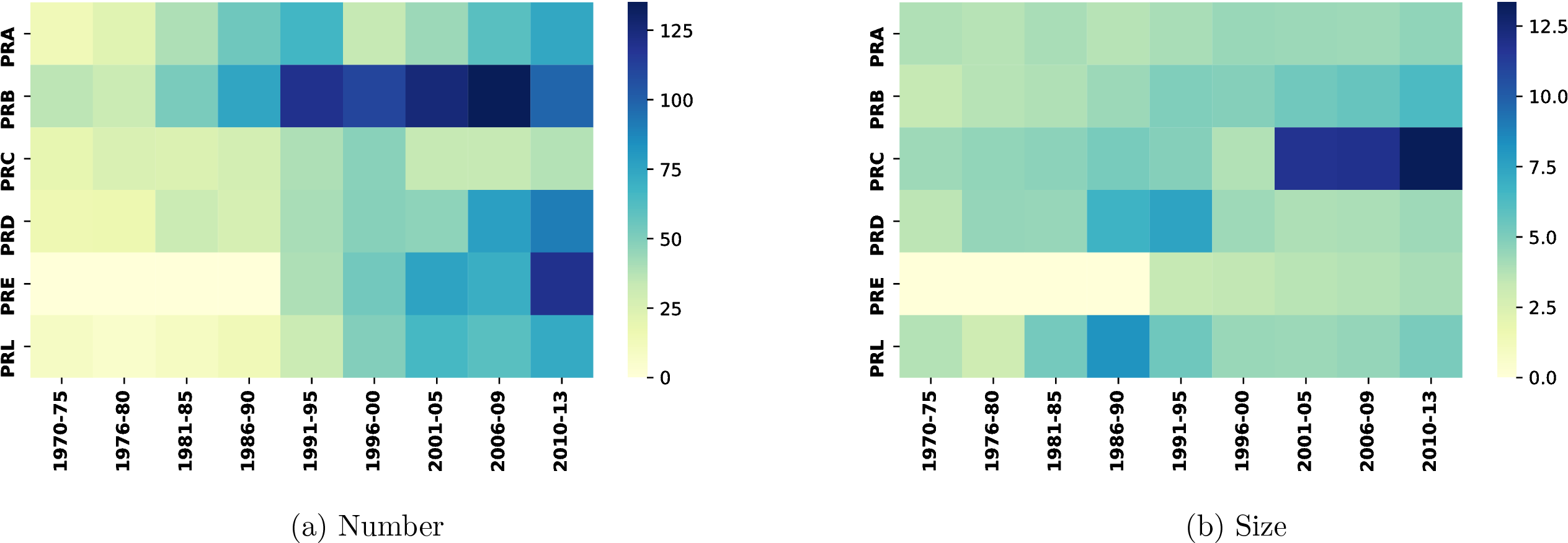}
	\caption{Growth in number of communities with time considering only Indian nodes. For PRE the initial values are zero because the journal started publishing in the 90's}
	\label{fig:gr_ci44}
\end{figure*}

There are many  methods to extract community structure in a network such as hierarchical clustering, spectral partitioning, modularity optimization and clique percolation method(CPM) \cite{fortunato2010community}. The co-authorship network is a clique dominated network
as authors collaborating in a paper are considered to be fully connected. Cliques can have overlap between communities as researcher can collaborate with different groups in a given time period. For our study we use CPM to extract the underlying community structure for evolving graphs,and Louvain method (based on modularity optimization) community detection for the union graph.  In Fig. $\ref{fig:gr_ci44}$ (a), we show the variation of number of communities for different APS journals. Here we have excluded all the foreign authors from the network to reveal the communities of Indian researchers. PRB has the most number of communities in almost all time periods (except 2010-13). All the other journals show a gradual rise in the number of communities with time. Fig. $\ref{fig:gr_ci44}$ (b) represents the average size of communities over time for all APS journals. Community in APS journals in the constructed network is mostly dominated by small sized communities (size 3-10). The maximum size for communities is of $O(10)$ for PRA,PRB and PRE while it is of $O(10^2)$ for PRC,PRC and PRL.  PRC has the largest average size 2001 onwards.  While PRD and PRL peak between 1986-1995. Network for all the other journals is mainly dominated by relatively smaller communities.

\begin{figure*}[ht]
	\centering
	\includegraphics[width=1.0\textwidth]{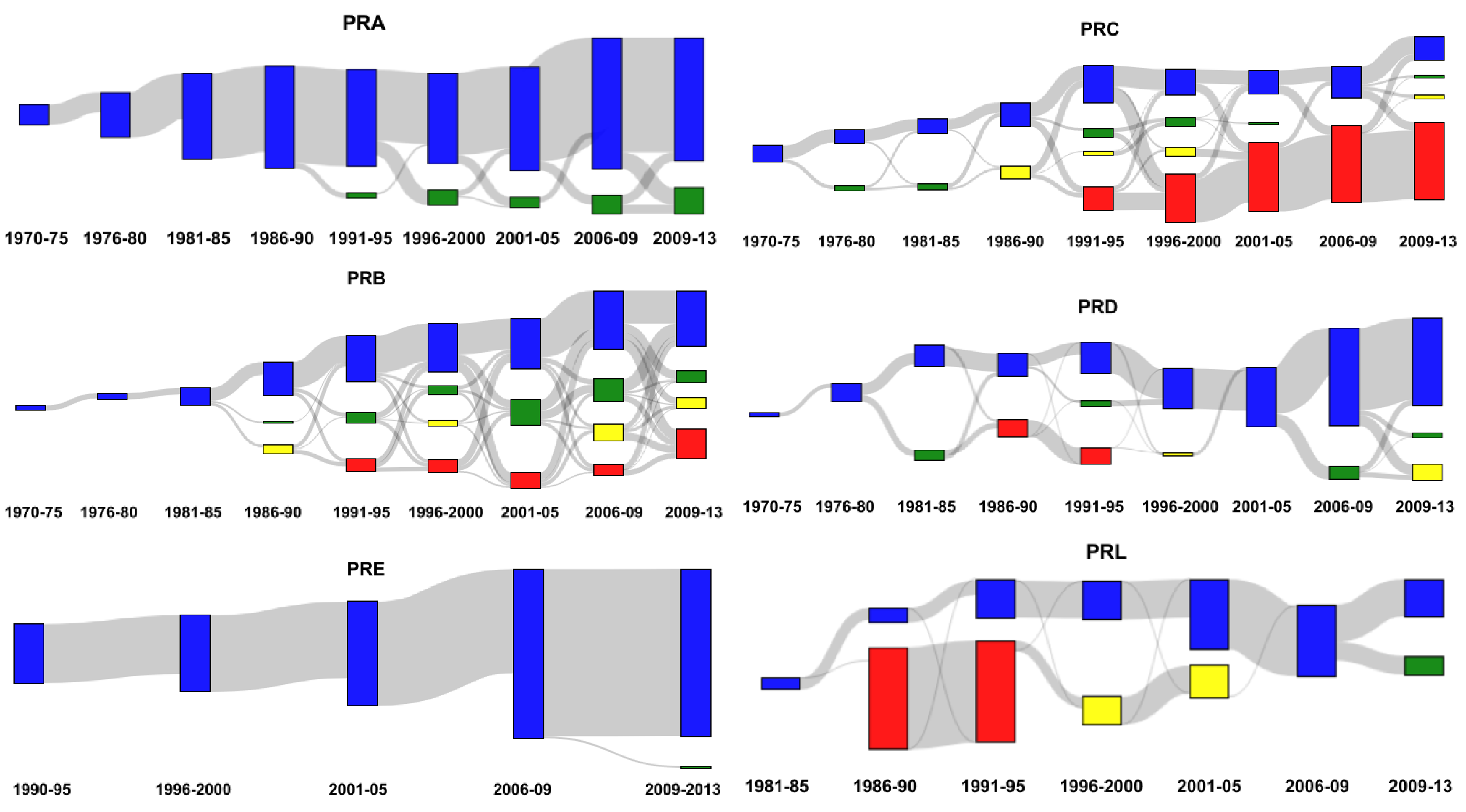}
	\caption{Flow of authors between communities of different sizes clubbed into 4 groups: the  \fcolorbox{black}{blue}{\rule{0pt}{6pt}\rule{6pt}{0pt}}\quad represents tiny(3-10),  \fcolorbox{black}{green}{\rule{0pt}{6pt}\rule{6pt}{0pt}}\quad small(11-20), 
		\fcolorbox{black}{yellow}{\rule{0pt}{6pt}\rule{6pt}{0pt}}\quad medium(21-35) and \fcolorbox{black}{red}{\rule{0pt}{6pt}\rule{6pt}{0pt}}\quad large(35+) community. Size of the bar represents the number of authors in communities of corresponding size.  }
	\label{fig:comm_dyn_flow1}
\end{figure*}

Analyzing the structure and evolution of communities can give us information about the growth pattern of the network.While centrality measures help us infer the role of individual in the network, community based analysis helps in understanding group dynamics. The latter gives better insight into the pattern of collaboration between individuals as groups. Community detection based analysis is done in three steps. First, we look at the overall community  structure of the network by extracting communities from the combined network of 43 years for all journals (except PRE for which the span was 20 years). Observations from this method are discussed in the subsection (structure of communities). Second, we study the communities network in  9 time periods between 1970-2013 and analyze the movement between communities of different sizes. Third, we break our data into one years intervals  and trace the life and stationarity of communities by size. To increase clarity we grouped communities into 4 major categories based on their sizes and map the flow of authors between these sizes based community groups shown in Fig \ref{fig:comm_dyn_flow1}. 

%\subsubsection{Structure of communities}

\begin{figure*}[ht]
	\centering
	\includegraphics[width=1.0\textwidth]{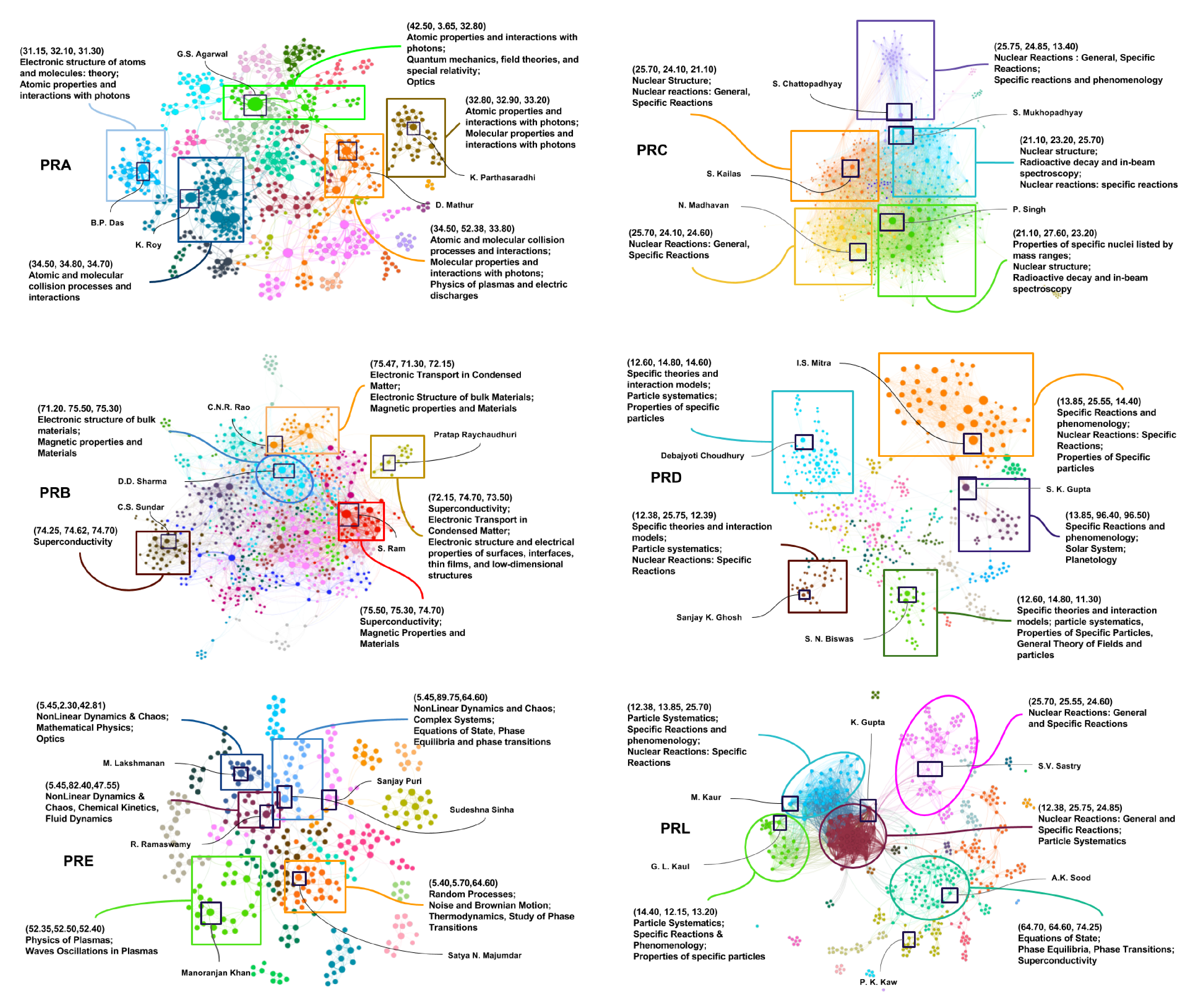}
	\caption{Indian Physics coauthorship network (1970-2013) in major APS journals  PRA, PRB,  PRE ;  PRC, PRD, and PRL, with major communities and their main field of research.}
	\label{fig:complete_dyn_1}
\end{figure*}

In Fig. \ref{fig:complete_dyn_1}, we show the static structure of Indian physics coauthorship network.  The APS started assigning PACS (Physics and Astronomy Classification Scheme) subject codes from 1980 onwards, which became a regular feature  from 1984 onwards. PACS codifies diverse research areas in different fields of physics.   To extract the major field of the community, we extract the most frequent PACS code of papers published by the members of the community (see supplemtary material for details and illustration). We highlight the top 3 most occurring PACS and their subjects for the communities, and also point out the person with the highest degree in these communities.

In Fig.\ref{fig:comm_dyn_flow1}, we  explore the migration of authors between different size communities, to get a broad idea about stability and inter community dynamics using alluvial flow diagrams \cite{vehlow2015visualizing,herrera2010mapping}. Communities were clubbed into four different size groups. The size range for the group were classified as following Tiny(3-10), Small(11-20), Medium(21-35) and Large(35+).  The dominance of tiny sized community groups indicate that a strong collaboration within small groups in PRA and PRE.  In PRB, we find that larger communities are being formed over time, and inter community transition happens across communities of all sizes.  In PRC, we notice an increased presence of large communities from 1991 onwards, and they dominate the dynamics. PRD has largely  tiny and small communities, but occasionally we do see medium and large size communities.
For PRL, only papers with single or two Indian authors were there from 1970-1980, hence we do see any community dynamics in this time period.  From 1986-90 and 91-95 has shows large communities in PRL. We have verified that this is due to large collaboration papers with foreign authors, who also had Indian affiliation. 

\begin{figure*}[ht]
	\centering
	\includegraphics[width=1.0\textwidth]{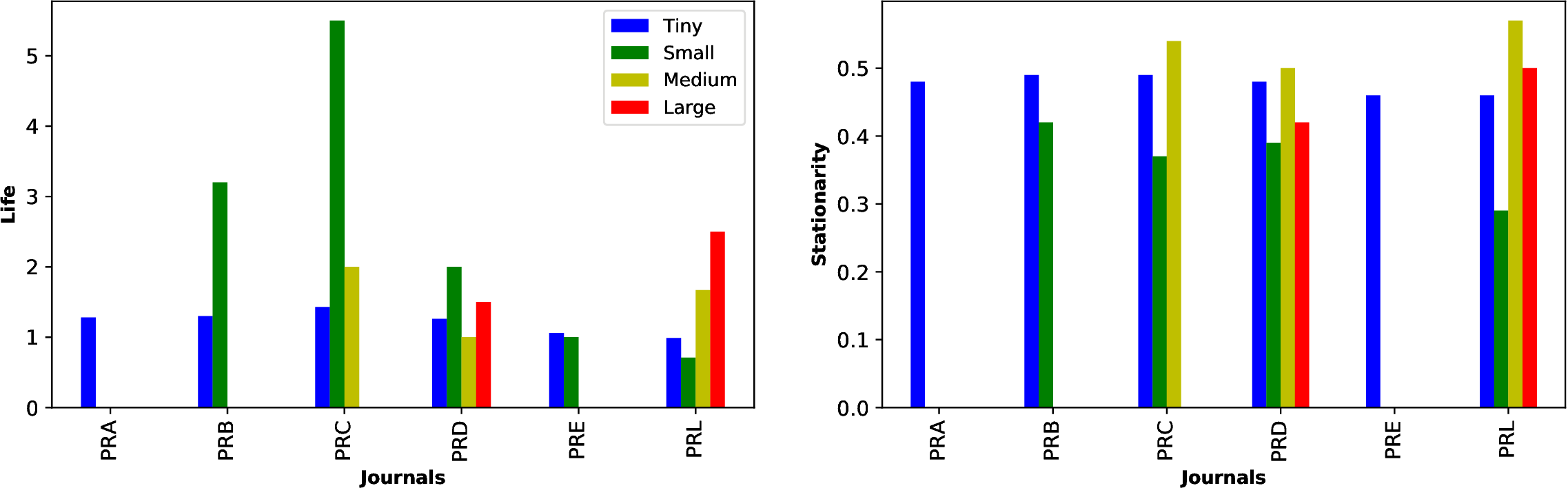}
	\caption{Average life time and stationarity $\langle \zeta \rangle$ of communities by size of Indian authored articles in APS journals. The classification of size is as shown in  the caption of \ref{fig:comm_dyn_flow1}  }
	\label{fig:life_size&st}
\end{figure*}

To characterize and trace the evolution of communities over time we use the measures for life time and stationarity from Palla \textit{et al.} ~\cite{palla2007quantifying, palla2009social} (see methods).  We divide the network yearly for all APS journals and extract the community structure using Clique Percolation Method (CPM) with k=3 \cite{fortunato2010community} year wise and track the growth, collapse, merger, split and death of communities \cite{palla2007quantifying}.  Fig $\ref{fig:life_size&st}$ shows the relationship between life, size and stationarity ($\zeta (A)$, defined in methods) of communities for various APS journals. We notice that mean life of communities in all groups is between 1-2 years except for PRB and PRC where medium size communities show a longer life.  In Fig \ref{fig:life_size&st},  plots the average stationarity ($\langle \zeta \rangle$) of communities for all APS journals is also shown. Tiny communities for all journals were observed to have stationarity of $\approx$ 0.48. The dynamicity ${\cal D} (A)$ of a community $A$, defined as ${\cal D} (A)=1-\zeta (A)$ measures the volatility in the members of communities.   I general small communities sized 11-20 are most dynamic. For PRC, PRD and PRL, the medium sized communities are least dynamic.  The dynamicity of communities can be related  to the appearance of new authors over time in APS journals publications (see Appendix information for more details).

\section*{Discussion}

To the best of our knowledge, this is the first study tracing the evolution of Indian physics 
spanning close to a century from 1919-2013. We have assessed the relative contribution of India to different fields of Physics through descriptive and network based analysis of Indian co-authorship network in American Physical Society journals. We have tracked the collaboration pattern of Indians with other Indians and Foreigners. We have characterized structural differences in co-authorship between different journals, specifically highlighting the differences between low energy and high energy physics journals. We have extracted community structure of Indian Physics networks and tracked the evolution of communities from their birth to death by size. A brief summary of highlights from the current work is given below.

 India has 1/6$^{th}.$ of the world population but its relative contribution to physics publications has been small at $3\%$ with significant variations across disciplines. Although maximum number of Indians have been contributing to condensed matter physics and material physics, relative contribution has been  higher in high energy physics journals PRC and PRD.  Indian co-authorship network has been growing steadily over the years, increasing collaboration with foreign authors. Growth rate of Indian collaboration with foreigners is higher than that with Indians within. Maximum growth in collaborations cab be seen in PRL. stands out among all the The mean active period for an Indian is around 4.5 years, with substantial variation among journals. We find strong structural difference between low energy physics (PRA,PRB and PRE) and high energy physics(PRC,PRD and PRL) coauthorship networks. The power law exponent for degree distribution of authors is between 2-3 for PRA, PRB and PRE , and between between 1-2 for high energy journals PRC, PRD and PRL, which is  mainly due to the presence of large collaboration groups in high energy journals. Such  differences can also been seen in other network measures such average degree $\langle k \rangle$ and average path length $\langle l\rangle$. For high energy journals, removal of foreign authors makes significant difference to the network measures, where as for PRA, B, and E, the differences are not significant.

Using overlap of ranking by different centrality measures we have extracted most influential authors in every time period for each APS journals. Though some authors show dominance in multiple time periods, by and large 5 years is period of maximum influence for Indians. This can also be compared with average active period for Indians. Longer active period is not directly linked dominance in the co-authorship network. The evolution of communities between 1970-2013 reveal the inherent dynamics between communities of different size groups. PRA and PRE have only tiny and small communities. PRB has largest number of communities, while PRC shows the largest size of community. Transition of authors  is largely between tiny to small sized communities, except for PRC. Stationarity analysis reveal that communities of Indian physicists are highly dynamic and do not live longer baring some exceptions in PRB and PRC.

%\begin{enumerate}
%	
%	\item Growth of number of communities. 
%	\item Major Communities and Significant members in each for all journals.
%	\item Observed mean Life and Stationarity of Communities foe all APS journals.
%	\item Community Dynamics changing over time.
%	\item Comment on the contribution and implication of this study. What can be done in future with this study.
%	\item New vs. Old authors. (figure is in Appendix)
%	\item Changing Affiliations with time. (figure is in Appendix)	
%\end{enumerate}

Our study highlights the significance of country specific network analysis in revealing the collaboration pattern between researchers. This study could be extended to study the pattern of collaborations by Geography and Institutions.  The structural characterization of co-authorship networks and their evolution can assist the science policy makers in prioritizing different research areas, building manpower and funding to enhance country's contributions to theoretical, experimental and applied research in Physical sciences. 

\section*{Methods}

%XXX ---------------------------- 
%
%Up to three levels of \textbf{subheading} are permitted. Subheadings should not be numbered.

%\section*{Discussion}
%
%The Discussion should be succinct and must not contain subheadings.

\textbf{~~~~Collaboration Coefficient} is defined as $CC=1-E[1/X]$, where $X$ is the number of authors in an article and $E[1/X]$ is the average credit awarded to each author \cite{ajiferuke1988collaborative} (assuming equal contributions). A sample estimate of CC can be obtained by:
\begin{equation}
CC = 1 - \frac{\sum_{j=1}^{q}\frac{1}{j}f_j}{N} \tag{1}, \label{eq:1}
\end{equation}
where $f_j$ number of $j$ authored papers published in a discipline in the period of study, $N$ is the total papers published in that time period, and $q$ represent the maximum no of authors in a paper in the discipline. 

\textbf{Co-authorship network construction:} We construct a bipartite graph of authors and papers for each journal at intervals of one and five years from 1970-2013. For PRC, PRD, and PRL,  we set a natural breaking point  (as described in main text) after 300 as a line in Fig. \ref{fig:c_auth} a, to eliminate edges in papers by large collaboration groups. From this graph, we form co-authorship networks with authors as nodes and weighted edges representing the frequency of co-authorship between two people. Based on the author's affiliation, we label each author as I (Indian) and Foreigner (F). The structure and characteristics of co-authorship networks were computed using python packages like networkx ~\cite{hagberg2005networkx} module, and visualization was primarily done through Gephi \cite{ICWSM09154}. 

\textbf{Tracking authors in networks over time:} While tracing the network evolution over time, we have to account for changing naming styles by the same author in different time periods. The differentiation between two authors is done by comparing names as strings. To identify the similarity between author's name we compare the neighborhood of authors in the network. Author's names with significant intersection are compared over on-line databases to confirm the uniqueness of identity. However we realize that for such a huge database there is still scope for some errors. But by large this resolves the error due to variation in naming style over time keeping it consistent and unique for authors, the network statistics and properties are also consistent.

\textbf{Centrality measures: } Degree centrality of author $i$  ($k_i=deg(i)$) represents the number of distinct collaborators of author in the network; Betweenness centrality ($b_i=\sum_{st} \frac{n^i_{st}}{g_{st}}$, $n^i_{st}$ is the number of shortest paths from $s$ to $t$ passing through node $i$, $g_{st}$ is the total number of shortest paths from $s$ to $t$) measures the extent to which an author lies on the shortest paths between other authors.  Closeness centrality of an author ($C_i=\frac{n}{\sum_j d_{ij}}$) measures the mean shortest path length between the author and all other nodes in the coauthorship network \cite{newman2010networks};  Neighborhood coreness centrality $C_{nc}$) ~\cite{bae2014identifying} is defined as $C_{nc}=\sum_{w\in N(v)}k_s(w)$ ,  where $k_s(w)$ is the k-shell index of node $w$, and $N (v)$ is the set of neighbors of node $v$.  

\textbf{Community life span and stationarity } Once a community is born, it can grow, merge , split or die in the next time interval. The life span of a community is the difference between the birth $t_0$, and the death $t_n$ , when $A(t_n)=0$.  The overlap of a community $A$  between time $t$ and $t+1$ is defined as \cite{palla2007quantifying, palla2009social} :
\[
C(A; t) = \frac{|A(t_0)\cap A(t_0 + t)|}{|A(t_0)\cup A(t_0 + t)|}. \tag{2} \label{eq:2}
\]
The stationarity $\zeta$ of a community $A$ during its lifespan is defined as:
\[
\zeta (A) \equiv \frac{\sum_{t=t_0}^{t_{max}-1}C(t,t+1)}{t_{max}- t_0 - 1} . \tag{3} \label{eq:3}
\]
\section*{Acknowledgements}
We acknowledge American Physical Society for providing us meta data of all APS publications till 2013. 

\section*{Author contributions statement}
C.S. conducted all the coding, data analysis and visualization and writing. S.J. conceptualized the problem, analyzed the results and supervised the writing.   All authors reviewed the manuscript. 

\section*{Additional information}

To include, in this order: \textbf{Accession codes} (where applicable); \textbf{Competing financial interests} (mandatory statement). 

The corresponding author is responsible for submitting a \href{http://www.nature.com/srep/policies/index.html#competing}{competing financial interests statement} on behalf of all authors of the paper. This statement must be included in the submitted article file.

%\noindent LaTeX formats citations and references automatically using the bibliography records in your .bib file, which you can edit via the project menu. Use the cite command for an inline citation, e.g.  \cite{Figueredo:2009dg}.

\section*{Appendix}

\section{Indian authors in APS papers}

\begin{figure*}[htbp]
	
	\centering
	\includegraphics[width=1.0\textwidth]{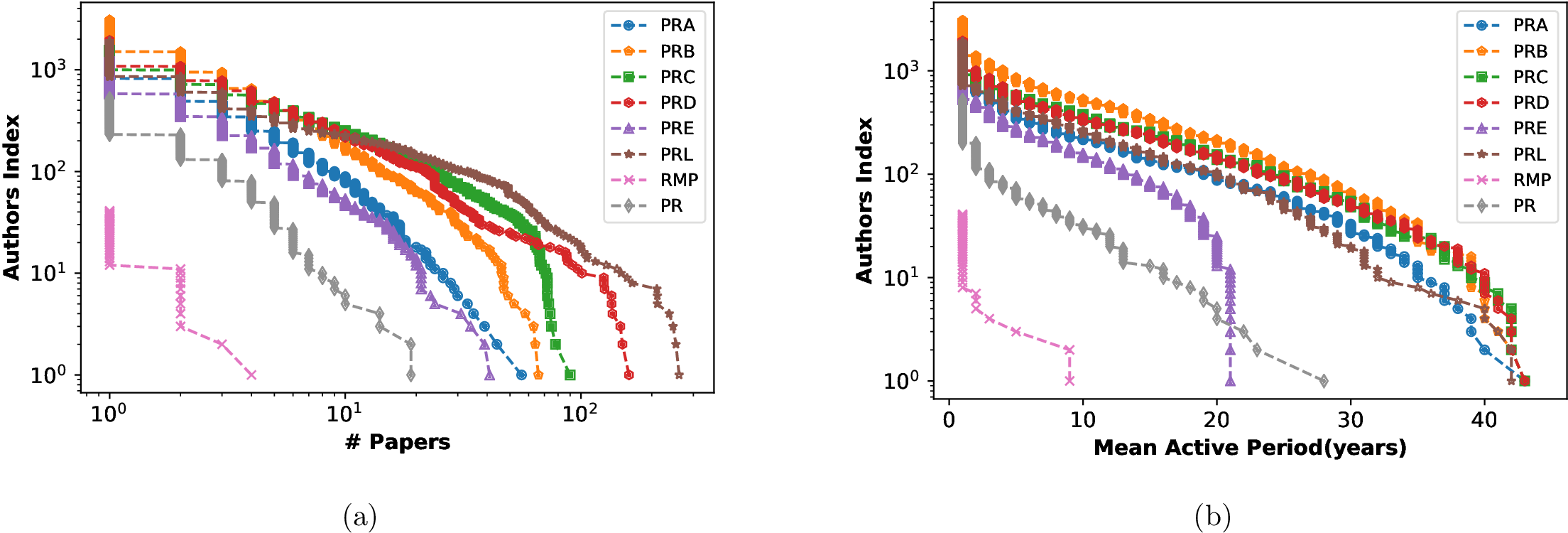}
	\captionsetup{singlelinecheck = false,justification=raggedright, font=footnotesize, labelsep=space}
	\caption{Panel (a) index of authors and the number of papers, panel (b)Mean active period of authrs . Active period (T) is defined as follows T = Year Last Published - Year First Published + 1, from 1919-2013. }
	\label{auth_pap_t}
\end{figure*}

The data studied in the main manuscript  spans almost a century (1914-2013) of publications having at least one Indian affiliations. A total of 14,704 papers were published having Indian affiliations. In Fig \ref{auth_pap_t} (a)  gives  index of authors and the number of their papers in APS publications with atleast one Indian author and  panel (b) shows the mean active period of Indian authors based on their publications in APS journals.  

\begin{figure*}[htbp]
	
	\centering
	\includegraphics[width=0.8\textwidth]{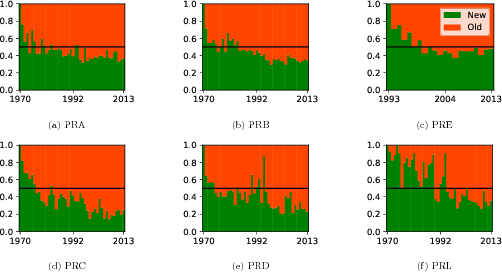}
	
	\captionsetup{singlelinecheck = false,justification=raggedright, font=footnotesize, labelsep=space}	
	
	\caption{Appearance of new Indian Authors in various APS journals with time.}
	\label{autn_n_o}
	
\end{figure*}

\section{Tracking the appearance of new authors}

Co-authoring papers is primarily a social phenomenon influenced by factors like field expertise of collaborators, personal connections,  supervisor-student interactions, and openness of a community towards a new entrant.  The ability of new entrant to publish in a given journal can be understood by looking at a fraction of new authors versus the whole in a given period. We compare the new vs old authors year wise in different journals in Fig \ref{autn_n_o}.  The PRA, PRB and PRE journals show the stable structure in the fraction of new authors up to 1990, and thereafter shows a  marginal decline. The PRL shows a downward trend with maximum fluctuations, while PRC and PRD show steeper decline with some fluctuations.

\section{Affiliation Dynamics}

\begin{figure*}[htbp]
	
	\centering
	\includegraphics[width=1.0\textwidth]{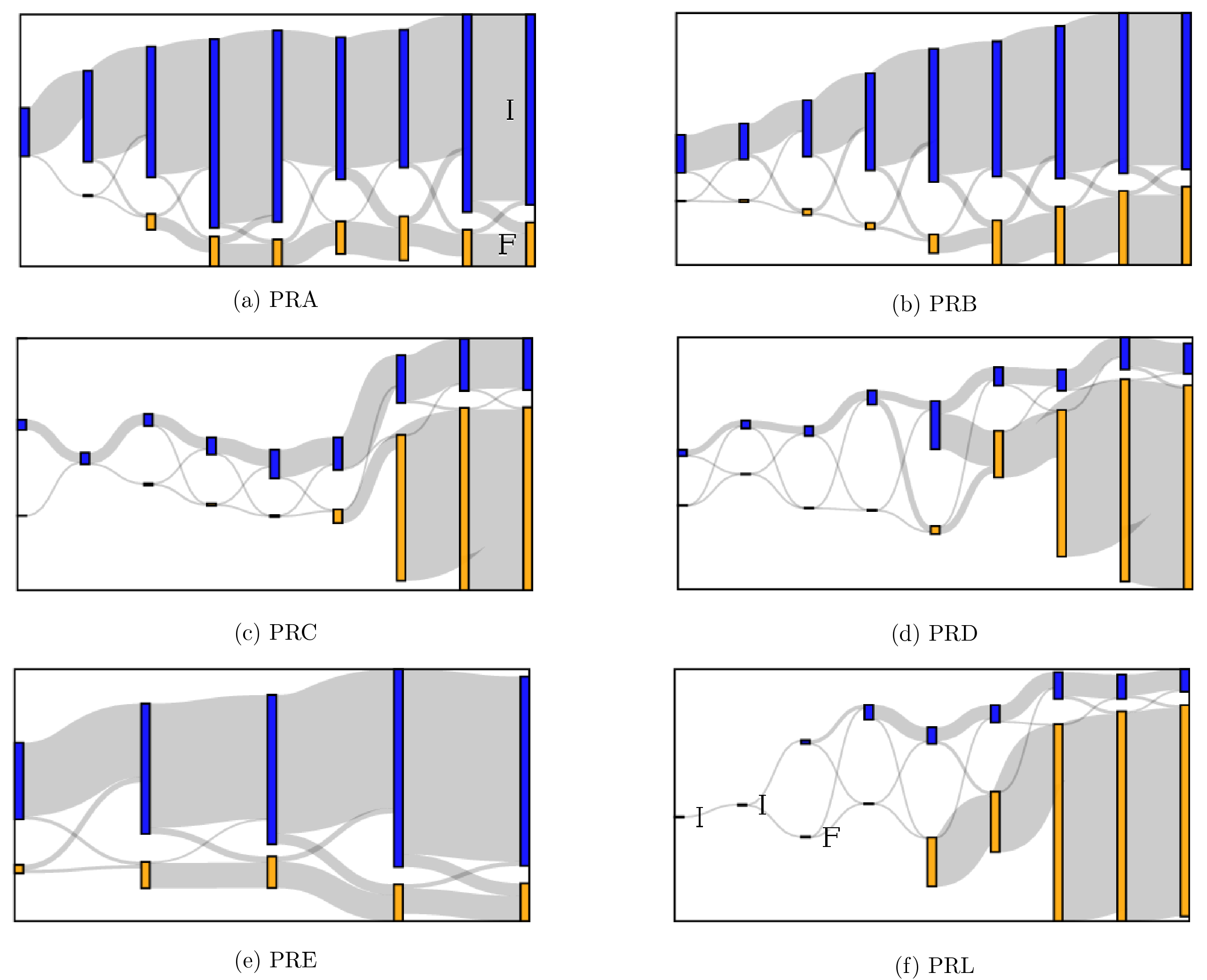}
	
	\captionsetup{singlelinecheck = false,justification=raggedright, font=footnotesize, labelsep=space}	
	
	\caption{Change in affiliation of authors with time separately for each APS journal. Blue represents Indian authors while Orange represents foreign authors.}
	\label{aff_Dyn}
	
\end{figure*}

Affiliations of authors keeps changing with time due to movement of scientists to different Institutions. We trace the mobility of Indian physicists from India to foreign and vice versa in five year intervals in Fig \ref{aff_Dyn}. Increase in mobility across institutes reflects the collaboration with different countries over time.

While tracing the network evolution over time, we have to account for changing naming styles by the same author in different time periods. The differentiation between two authors is done by comparing names as strings. To identify the similarity between author's name we compare the neighborhood of authors in the network. Author's names with significant intersection are compared over on-line databases to confirm the uniqueness of identity. However we realize that for such a huge database there is still scope for some errors. But by large this resolves the error due to variation in naming style over time keeping it consistent and unique for authors, the network statistics and properties are also consistent.

\section{Community structure}

\begin{figure*}[htbp]
	
	\centering
	\includegraphics[width=1.0\textwidth]{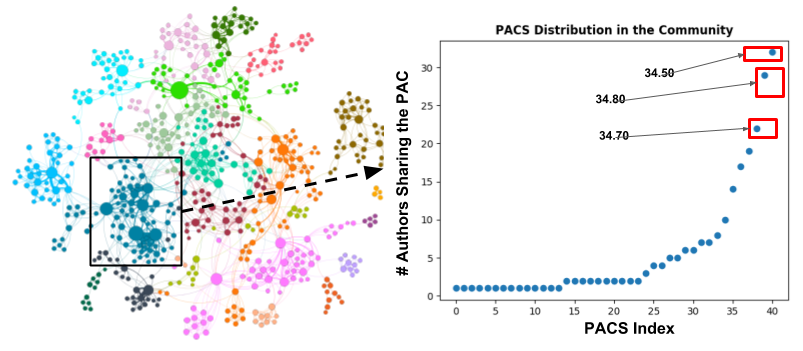}
	
	\captionsetup{singlelinecheck = false,justification=raggedright, font=footnotesize, labelsep=space}	
	
	\caption{To identify the major research interests of a community, we extract top 3 most occurring common PACS for all members in the community.}
	\label{cmNpc}
	
\end{figure*}

To extract major research groups in Indian Physics co-authorship network, we consider union graphs constructed for each journals. Louvain method was used for community extraction after removal of Foreign nodes. In order to identify the research interest of the community members we extract the publication data for all authors in the community. Top 3 most occurring PACS up to second level of classification were identified as the groups field of interest as shown in Fig \ref{cmNpc}.

\bibliography{main_p}

\end{document}